\documentclass[12pt]{iopart}

\usepackage{graphicx}
\usepackage{dcolumn}
\usepackage[justification=justified,singlelinecheck=false]{caption}
\usepackage{bm}

\makeatletter

\newcommand{\Rmnum}[1]{\expandafter\@slowromancap\romannumeral #1@}
\makeatother

\begin{document}

%\preprint{}

\title{A study of transport suppression in an undoped AlGaAs/GaAs quantum dot single-electron transistor}



\author{A. M. See, O. Klochan, A. P. Micolich, M. Aagesen, P. E. Lindelof and A. R. Hamilton }

%\affiliation{School of Physics, University of New South Wales,
%Sydney NSW 2052, Australia}

%\affiliation{Nanoscience center, University of Copenhagen,
%Universitetsparken 5, DK-2100 Copenhagen, Denmark}

%\email{Alex.Hamilton@unsw.edu.au}
%\affiliation{School of Physics, University of New South Wales,
%Sydney NSW 2052, Australia}

%\thanks{}


\date{\today}

\begin{abstract}
We report a study of transport blockade features in a quantum dot single-electron transistor, based on an undoped AlGaAs/GaAs heterostructure. We observe suppression of transport through the ground state of the dot, as well as negative differential conductance at finite source-drain bias. The temperature and magnetic field dependence of these features indicate the couplings between the leads and the quantum dot states are suppressed. We attribute this to two possible mechanisms: spin effects which determine whether a particular charge transition is allowed based on the change in total spin, and the interference effects that arise from coherent tunneling of electrons in the dot.
\end{abstract}

%\keywords{single-electron transistor, undoped heterostructure, spin blockade, negative differential conductance, ground state suppression}

\maketitle

Semiconductor quantum dots have been used to realize single-electron transistors (SETs) \cite{Meirav:1990prl}, artificial atoms \cite{Kastner:1993pt}, ultrasensitive electrometers \cite{Fujisawa:2004apl} as well as elements for quantum information applications \cite{Hanson:2007rmp}. In particular,  the combination of spintronics and quantum information processing in semiconductor quantum dots has recently attracted considerable interest \cite{Loss:1998pra,Recher:2000prl,Hu:2001pra}. In such systems, studies of the interactions between spin and orbital states are of great importance. We have recently demonstrated a quantum dot single-electron transistor (Qdot-SET) based on an undoped AlGaAs/GaAs heterostructure without modulation doping \cite{See:2010apl}. Undoped devices are attractive due to their excellent charge stability \cite{Klochan:2006apl} and thermal robustness \cite{See:2012prl}. Indeed, we have shown that in an undoped open quantum dot, the magneto-conductance fluctuations, which are representative of the dot's overall potential (including disorder), are reproducible even after thermal cycling to room temperature \cite{See:2012prl}. Given these attractive properties of undoped devices, the natural question arises: can they be used to measure spin-related transport effects? In this paper, we address this question by studying transport blockade features in a Coulomb blockaded quantum dot, which can be explained by spin related effects.\\

\begin{figure}[h]
\centering
\includegraphics[width=9cm]{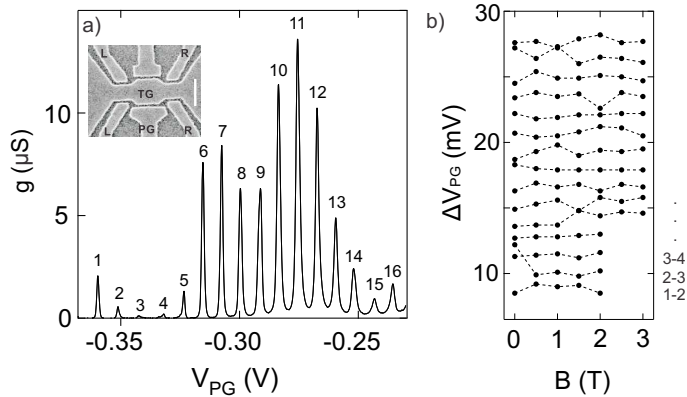}

\caption[width=10cm]{(a) Coulomb blockade oscillations are shown as two-terminal conductance $g$ vs. plunger gate voltage $V_{PG}$, obtained at $V_{L} = 0$~V, $V_{R} = -70$~mV and $V_{TG} = 0.85$~V.  A SEM image of the device is shown as an inset with a 500~nm scale bar. In (b), the spacing between adjacent CB peaks is plotted versus the magnetic field $B$, applied perpendicular to the 2D plane. The bottom trace corresponds to the spacing between peak 1 and peak 2. Traces are offset vertically by a constant $1.4$~mV to account for the charging energy and more clearly demonstrate the $B$ dependence of the dot levels.}
\end{figure}

A scanning electron microscope image of our AlGaAs/GaAs undoped quantum dot device is shown in the inset of Fig. 1. The two-dimensional electron gas is located 185 nm beneath a degenerately n$^{+}$ doped cap layer. A quantum dot with dimensions $0.54 \times 0.47~\mu$m was defined by etching the cap layer into seven separate gates using electron beam lithography and chemical wet etching. Based on the single particle energy level spacing of $\sim 210~\mu$V \cite{See:2010apl}, our dot contains about 60 electrons. Measurements were performed using standard dc techniques and ac lock-in techniques in a two-terminal configuration in a dilution fridge with an electron temperature of about $140$~mK \cite{See:2010apl}. Figure 1(a) shows a set of Coulomb blockade (CB) oscillations. To demonstrate that our dot operates in the quantum regime where well-defined single particle energy levels can be resolved, we examine the CB peak spacing $\Delta V_{PG}$, which is proportional to the addition energy. In Fig. 1(b), $\Delta V_{PG}$ versus the magnetic field $B$ applied perpendicular to the 2DEG is shown, where the bottom trace corresponds to the separation in $V_{PG}$ between peak 1 and peak 2 (P1 and P2). In general, the single particle energy levels in a dot split and can cross each other when $B$ couples to the orbital and spin parts of the dot's eignstates; for a parabolic confinement potential, these splittings result in the well known Fock-Darwin spectrum \cite{Fock:1928zp, Tarucha:2000apa, Ciorga:2002prl}. Although our dot is not parabolic, we do observe level crossings in Fig. 1(b), consistent with the results presented in Ref \cite{See:2010apl}, where the dot was found to operate in the quantum regime.\\

\begin{figure}[h]
\centering
\includegraphics[width=9cm]{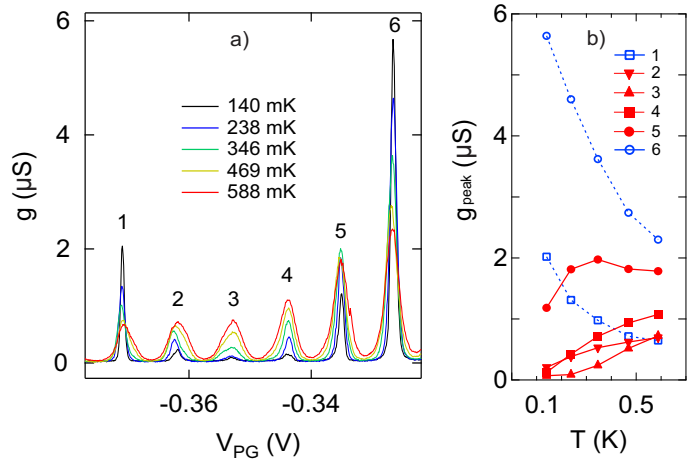}
\caption{(a) Temperature dependence of the Coulomb blockade oscillations for P1 to P6 in Fig. 1(a). (b) The peak conductance $g_{peak}$ is plotted against $T$. As expected for peaks in the quantum regime, $g_{peak}$ of P1 and P6 (blue dashed lines) decrease as $T$ is increased. In contrast, $g_{peak}$ for P2 to P5 (red solid lines) show an increase in conductance as $T$ increases.}
\end{figure}

We now focus on CB peaks P1 to P6, of which P2 to P5 are considerably smaller than the other peaks in Fig. 1(a). In order to understand this suppression, we perform a temperature dependence analysis,  as shown in Fig. 2(a). In Fig. 2(b) we plot the peak conductance $g_{peak}$ of the six conductance peaks versus temperature. As $T$ is increased, $g_{peak}$ of P1 and P6 decrease as expected for a dot in the quantum regime \cite{Kouwenhoven:1991prb}. However for P2 to P5, there is an anomalous increase in $g_{peak}$ as $T$ goes up \cite{Nicholls:1993prb}. This behaviour is not expected for a dot in the classical Coulomb blockade regime where $g_{peak}$ is not temperature dependent, nor in the quantum regime when transport is via the ground state \cite{remarks}.\\

\begin{figure}[h]
\centering
\includegraphics[width=9cm]{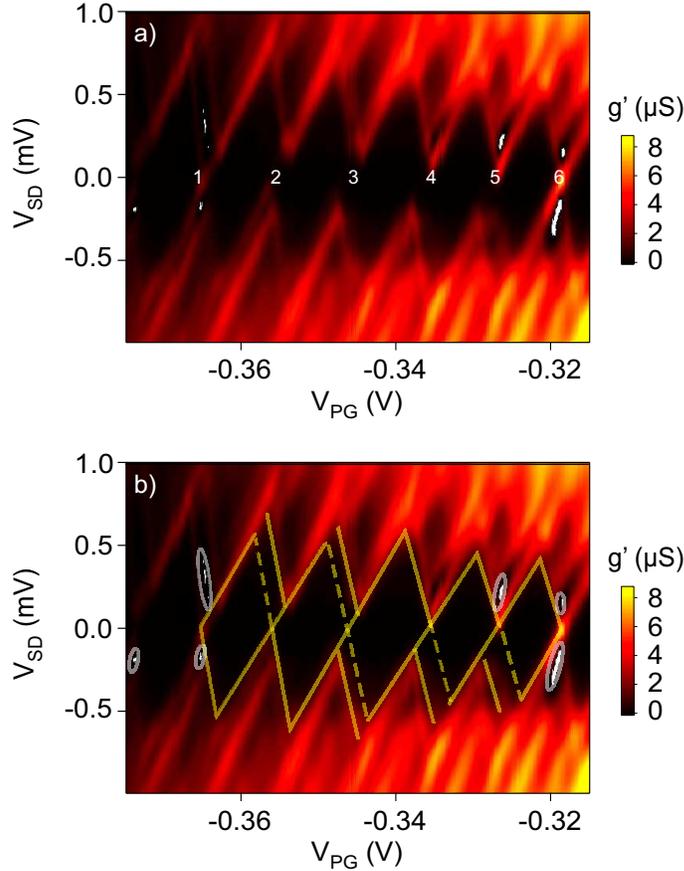}
\caption{(a) Bias spectroscopy of the Coulomb blockades peaks in Fig. 2(a), showing differential conductance $g^{\prime}$ (colour axis) as a function of plunger gate voltage $V_{PG}$ (x-axis) and dc source-drain bias $V_{SD}$ (y-axis). Zero bias conductance suppressions for P2 to P5 appear as “gaps” in the Coulomb diamonds. (b) The same stability diagram in (a) is shown with suppressed ground states highlighted by yellow dashed lines. Regions of negative differential conductance (NDC) for $g^{\prime} < -0.1~\mu$S are shown in white, and enclosed by white ellipses.}
\end{figure}

Further information on the nature of the anomalous peaks P2 to P5 is provided by performing bias spectroscopy on the Coulomb blockade peaks. In Fig. 3, the differential conductance $g^{\prime}$ (color axis) is plotted against the source-drain bias $V_{SD}$, and the plunger gate voltage $V_{PG}$. Dark regions indicate low $g^{\prime}$, and form a sequence of Coulomb diamonds where current through the dot is blockaded.
However for peaks P2 to P5, the suppressed zero bias conductance results in gaps in the diamonds at $V_{SD} = 0$~V. Transport is only allowed via the excited states for $|V_{SD}|> 0.1$~meV. These gaps are highlighted in Fig. 3(b), with the suppressed ground state transitions indicated by dashed lines. Both the temperature dependence analysis and bias spectroscopy of P2 to P5 suggest that conduction via the ground state is suppressed, and transport is re-enabled only when the an excited state becomes accessible via temperature activation or by increasing the source-drain bias. Interestingly we also observe regions of negative differential conductance (NDC) \cite{Johnson:1992prl,Weis:1992prb,Weis:1993prl}, highlighted by the white ellipses in Fig. 3(b), near peaks P1 and P6. The observed conductance suppression at $V_{SD} = 0$~V, as well as the NDC can be understood based on the differences in the coupling between the leads to the ground state, and to the excited state of the dot, as we now show. \\

\begin{figure}[h]
\centering
\includegraphics[width=9cm]{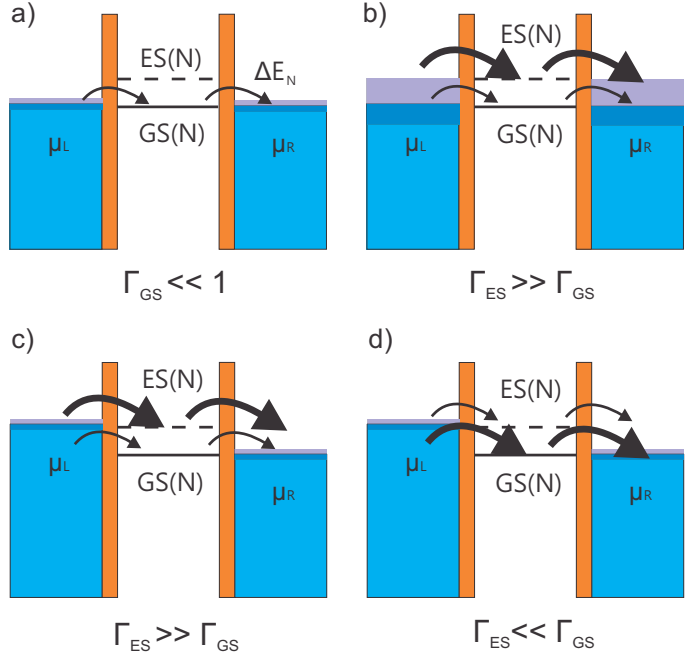}
\caption{These schematics illustrate transport blockade features based on the coupling rates between the leads and dot states. (a) The coupling between the leads and the ground state is anomalously low $\Gamma_{GS} \ll 1$, electron transport via the ground state is suppressed. Transport is allowed when a nearby excited state ES with $\Gamma_{ES} \gg \Gamma_{GS}$ becomes energetically accessible by means of (b) temperature activation, or (c) by increasing the source-drain bias. (d) In the case where $\Gamma_{ES} \ll \Gamma_{GS}$, $g$ is reduced when the excited state becomes accessible. Regions of negative differential conductance are expected to appear in the corresponding stability diagram.}
\end{figure}

In the sequential tunneling model of Qdot-SET transport, the current causes the ground state (GS) of the dot to oscillate between $N$ and $N+1$ electrons. This current is proportional to the coupling $\Gamma_{GS}$ between the leads and the dot's ground state \cite{Bonet:2002prb}. Transport through the ground state is suppressed when $\Gamma_{GS}$ becomes sufficiently low (see Fig. 4a), and hence the corresponding conductance peak will be significantly reduced. However, transport is re-enabled when a nearby excited state (ES), with a much larger coupling rate $\Gamma_{ES}$ to the leads compared to $\Gamma_{GS}$, becomes energetically accessible. Consistent with our observations in Figs. 2 and 3, this can be achieved by thermal activation, or by increasing the source-drain bias (see Figs. 4b \& c). In some cases, this suppression can also be lifted by applying a magnetic field $B$ \cite{Huttel:2003epl}. Conversely if the leads are coupled more strongly to the ground state than the excited state $\Gamma_{GS} \gg \Gamma_{ES}$ (see Fig. 4d), then $g$ will be reduced when the excited state becomes energetically accessible. This reduction in $g$ causes NDC as seen in Fig. 3.\\

One explanation for $\Gamma_{GS}$ being smaller than $\Gamma_{ES}$ could be due to fluctuations in the density of state (DOS) in the leads arising from disorder, as has been observed in the leads of self-assembled nanowire quantum dots and silicon MOS devices \cite{Bjork:2001nanolett,Takeuchi:1991prb,Matsuoka:1994jap,Mottonen:2010prb}. However in our case it is highly unlikely that the observed ground state transport suppression phenomena could be due to fluctuations in density of states in the leads, since the electron transport is ballistic with a mean free path of $\approx 2.1~\mu m$ and the leads are relatively wide  $(\approx 200~nm$) so that multiple 1D subbands are occupied. \\

Instead of disorder, there are two more likely transport suppression mechanisms which explain our data qualitatively. The first one relates to the total spin $S$ of the dot. Weinmann \emph{et al} performed an analysis of spin blockade effects in the linear and non-linear transport through a quantum dot and showed that ground state transport suppression and negative differential conductance at finite bias can be explained based on two sets of spin selection rules ~\cite{Weinmann:1995prl}. These rules were labelled as spin blockade type-\Rmnum{1} and type-\Rmnum{2}, which we briefly summarise here. \\

Type-\Rmnum{2} spin blockade suppresses the linear conductance, and occurs in transitions where the change in the total spin of the $N$ and $N-1$ electron ground states of the dot differ by more than 1/2:

\begin{equation}
GS(N, S) \leftrightarrow GS (N-1, S^{\prime}), \\
|S- S^{\prime}| > 1/2
\end{equation}

In this case, the transition between the $N$ and the $N-1$ electron ground state is spin blockaded, and the corresponding low bias conductance peak would be absent at $T = 0$~K. For example, if $S = 3/2$ for $N=5$ and $S^{\prime} = 0$ for $N = 4$, ground state transport is suppressed. However, transport can be re-enabled by accessing an excited state via means of temperature activation, or increasing the source-drain bias. Type-\Rmnum{2} spin blockade results in suppressed linear conductance, as seen in Fig. 1, and the transport recovery mechanisms are qualitatively consistent with the data presented in Figs. 2 and 3 respectively, where increasing $T$ or $V_{SD}$ increases $g$.\\

Type-\Rmnum{1} spin blockade relates to the population of states with maximal  $S = N/2$; the transitions:

\begin{equation}
(N, S=N/2)\to (N-1, S^{\prime}=(N-1)/2)
\end{equation}
which decrease the electron number from $N$ to $N-1$ and start with a spin polarized state $(S = S_{max}= N/2)$, will always reduce $S$ to $S^{\prime} = (N-1)/2$. In contrast, non-polarized states with $S < N/2$ can either increase or decrease the total spin after transition. Therefore the $S = N/2$ state is stable for relatively long time and as a result, when a spin-polarized excited state becomes accessible, the coupling between the leads and the dot's excited state is smaller than the coupling between the leads and the dot's ground state $\Gamma_{ES} \ll \Gamma_{GS}$. Signatures of type-\Rmnum{1} spin blockade are observed as regions of negative differential conductance, as seen in Fig. 3.\\

\begin{figure}[h]
\centering
\includegraphics[width=9cm]{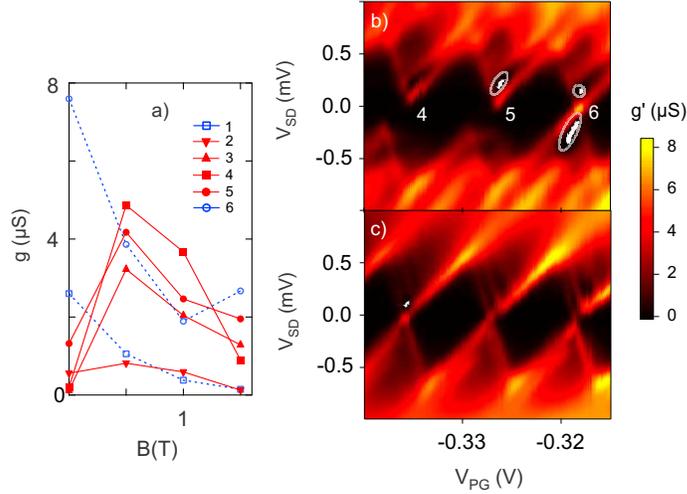}
\caption{(a) Coulomb blockade peak conductance $g_{peak}$ as a function of applied magnetic field $B$. The ground state conductance suppression for peaks P2 to P5 (red solid lines) is lifted at $B = 0.5$~T. (b) A stability diagram for P4 to P6 at $B = 0$~T similar to the one in Fig. 3(a) is shown. The ground state suppression features at P4 and P5 appear as “gaps” in the diagram, whereas regions of NDC for $g^{\prime} < -0.1~\mu$ S are shown in white. These spin-blockade features are lifted at $B = 1$~T and the corresponding stability diagram is shown in (c) The suppression gaps appear to be closed, and regions of NDC observed in (b) have disappeared in (c). }
\end{figure}

The other possible suppression mechanism relates to the coherent resonant tunneling of electrons inside the dot. To the first order approximation, a Qdot-SET is analogous to a Fabry-Perot resonator where the amplitudes of the Coulomb blockade oscillations are modulated based on the interference conditions \cite{Heinzel:1994epl}. Regions of suppressed Coloumb peaks can therefore be understood as a result of  destructive interference. These effects can be lifted via thermal broadening, where the degree of electron coherence is reduced. This behaviour is qualitatively consistent with our data in Fig. 2.   Similar argument can also be applied to our source-drain biasing data in Fig. 3, where the bias energy changes the interference condition by introducing a phase shift, or by reducing the electron coherence, thereby enabling transport via the excited states. Regions of negative differential conductance can also be understood as a result of constructive interference condition being lifted by the applied bias.\\

To test if the suppressed ground state conductance, and the NDC that we observe are spin related, we examine the magnetic field dependence of the Coulomb blockade peaks. In Fig. 5(a), a plot of $g_{peak}$ vs $B$ is shown. The conductance of peaks P2 to P5 which were strongly suppressed, rapidly increase with $B$, and then gradually decrease for $B > 0.5~$T. At higher $B$, the overall conductance through the quantum dot decreases because of a gradual compression of the dots states by the magnetic field \cite{Foxman:1993prb}. In contrast the amplitudes of peaks P1 and P6 generally decrease as $B$ is raised. This $B$ field dependence mimics the temperature dependence: there are two kinds of behaviour, with P1 and P6 showing different magnetic field dependence from P2 to P5, just as they showed different temperature dependence in Fig. 2.\\

In Fig. 5(b \& c) we compare the Coulomb diamonds that correspond to P4 to P6 at $B = 0$~T and $1$~T. The gaps near $V_{SD} = 0$~V indicate ground state suppression consistent with the type-II spin blockade identified in Ref ~\cite{Weinmann:1995prl}. When $B = 1$~T is applied, ground state transport is re-enabled, and the gaps disappear. Qualitatively, the dot levels get re-arranged under the influence of $B$, as shown in Fig. 1(b), which can lift the type-II spin blockade condition. The magnetic field also eliminates the regions of negative differential conductance (highlighted by white ellipses in Fig. 5(b)), as shown in Fig. 5(c). This can be similarly understood as the excited state becomes non-polarized due to level crossing at $B = 1$~T. However, as both the orbital and the spin states are affected by $B$ applied perpendicular to the 2DEG plane, a study of parallel field dependence of level shifting is required to unambigouosly identify spin blockade as the suppression mechanism responsible \cite{Weis:1993prl}. Nevertheless, our analysis is completely consistent with Ref \cite{Huttel:2003epl,Higginbotham:2012mm}, which reports the observation of spin blockade effects in a modulation-doped AlGaAs/GaAs Qdot-SET quantum dot device. \\

In summary, we have reported transport measurements of an undoped AlGaAs/GaAs quantum dot single-electron transistor where ground state transport suppression, as well as negative differential conductance were observed. These transport blockade features can be explained based on the relative couplings between the leads and discrete levels in the dot and we attribute the suppression mechanism to two set of spin selection rules proposed by Weinmann, and the coherent resonant tunneling of electrons. It might be argued that since our dot is large, and contains at least 60 electrons, spin dependent transport should not be observed. However our observations are consistent with  previous experiments on large quantum dots containing $\sim 50$ and $\sim 10$ electrons, where spin related effects were also observed \cite{Huttel:2003epl,Higginbotham:2012mm}. We thank Mark Eriksson, Klaus Ensslin and Sven Rogge for helpful discussions. This work was supported by the Australian Research Council Grants (ARC) [DP0772946, DP120101859, FT0990285]. A.R.H. acknowledges an ARC Outstanding Researcher Award.\\

\end{document}